\documentclass[prl,twocolumn]{revtex4}

\usepackage{amssymb,amsthm}
\usepackage{mathtools}

\newcommand{\C}{{\mathbb C}}

\newcommand\be{\begin{eqnarray}}
\newcommand\ee{\end{eqnarray}}

\newtheorem{proposition}{Proposition}

\begin{document}

\title{New first order Lagrangian for General Relativity}
\author{Yannick Herfray and Kirill Krasnov} 
\affiliation{School of Mathematical Sciences, University of Nottingham, UK}

\begin{abstract} We describe a new BF-type first-order in derivatives Lagrangian for General Relativity. The Lagrangian depends on a connection field as well as a Lie-algebra valued two-form field, with no other fields present. There are two free parameters, which translate into the cosmological constant and the coefficient in front of a topological term. When one of the parameters is set to zero, the theory becomes topological. When the other parameter is zero, the theory reduces to the (anti-) self-dual gravity. Thus, our new Lagrangian interpolates between the topological and anti-self-dual gravities. It also interprets GR as the (anti-) self-dual gravity with an extra quadratic in the auxiliary two-form field term added to the Lagrangian, precisely paralleling the situation in Yang-Mills theory.
\end{abstract}

\date{March 2015}
\maketitle

As years went by, Plebanski's formulation \cite{Plebanski:1977zz} of General Relativity has become more and more appreciated as an important insight into the structure of the theory. This is a first order formulation that has an ${\rm SO}(3,\C)$ connection and ${\mathfrak so}(3,\C)$-valued two-form field as the basic variables. The complexified fields are used to describe complexified General Relativity. Real metrics of various signatures are obtained by imposing appropriate reality conditions. In addition, the Plebanski Lagrangian contains a certain symmetric matrix-valued Lagrange multiplier field, which is tracefree. 

It was understood a decade later \cite{Capovilla:1989ac} that two of the basic fields of this formulation, namely the two-form field and the Lagrange multiplier field, can be "integrated out" from the Lagrangian, resulting in what can be called the "pure connection" formulation. In reference \cite{Capovilla:1989ac} this was described for the case of zero cosmological constant. In this case, the resulting "pure connection" description is not as pure as one might desire - the Lagrangian contains an additional (densitiesed scalar) auxiliary field. It was later realised in \cite{Krasnov:2011pp} that things work much nicer when there is a non-zero cosmological constant - in this case one gets a true "pure connection" formulation. In both cases non-polynomial constructs involving the square root (of a matrix) appear. 

Another related work is \cite{Krasnov:2006du}. This paper described a family of what can be called "deformations" of GR. These are obtained by making the cosmological constant in Plebanski's action an arbitrary function of the Lagrange multiplier field. These theories still propagate just two polarisations of the graviton, see e.g. \cite{Krasnov:2008zz}. Once the cosmological constant is replaced with a function of the Lagrange multipliers, one can integrate out the field of multipliers altogether (this is not possible in the case of GR without also integrating out the two-form field). This results in a class of theories that can be succinctly described by the following Lagrangians
\be\label{type}
{\cal L}[B,A] = B^i \wedge F^i + V(B^i\wedge B^j).
\ee
Here $A^i$ is an ${\rm SO}(3,\C)$ connection, $F^i$ is its curvature, $i=1,2,3$ is the ${\mathfrak su}(2,\C)$ Lie algebra index, and $B^i$ is the Lie algebra valued two-form. The function $V$ is a gauge invariant, homogeneous degree one function of symmetric $3\times 3$ matrices. This allows it to be applied to the 4-form $B^i\wedge B^j$, with the result being a well-defined 4-form. 

The aim of this letter is to point out that usual (undeformed) General Relativity (in 4 space-time dimensions) can be described by a Lagrangian of the form (\ref{type}). This should be contrasted with Plebanski's formulation, which is, as (\ref{type}), a first order one, but contains, in addition to $B,A$, a Lagrange multiplier field. Such a field is not present in (\ref{type}), and so one might prefer such Lagrangians as more economical. Let us also emphasise that there is no simple way to get rid of the Lagrange multiplier field in Plebanski's formulation, as the variation with respect to it imposes what is usually called the "metricity" condition $B^i\wedge B^j\sim \delta^{ij}$.

Let us now state our new Lagrangian. It depends on two parameters $\alpha, \lambda$ and reads
\begin{multline}\label{L}
{\cal L}_{\rm GR}[B,A] \\   = B^i\wedge F^i  +\frac{\alpha}{2} \left( {\rm Tr}(\sqrt{B^i\wedge B^j})\right)^2 - \frac{\lambda}{2} B^i\wedge B^i  .
\end{multline}
The first and the last terms here describe what can be called the BF theory with the cosmological term. Thus, when $\alpha$ is set to zero, the theory is a known and well-studied topological field theory. The new term is one with $\alpha$ in front. Here we encounter the already familiar from \cite{Krasnov:2011pp} structure of the trace of the square root of the matrix of wedge products. 

Our claim is that for generic values of $\lambda,\alpha$ the Lagrangian (\ref{L}) describes General Relativity with a non-zero cosmological constant, see below for the demonstration. A certain combination of the parameters $\lambda, \alpha$ plays the role of the cosmological constant. A certain other combination is a parameter in front of a topological term that can always be added to GR without changing the dynamics (analog of the $\theta$-term in QCD). The Newton's constant is an overall coefficient that is set to unity for simplicity. 

Perhaps the most interesting aspect of the new description (\ref{L}) is that in the limit $\lambda\to 0$ the character of the theory changes. As we show below, in this case (\ref{L}) describes nothing else but the (anti-) self-dual gravity. This theory propagates just one of the two polarisations of the graviton, and describes what is often called gravitational instantons. 

To explain why such a Lagrangian description of the gravitational instantons is an exciting development, we remind the reader situation in Yang-Mills theory. In that case, a possible first order formulation of YM is by the Lagrangian
\be\label{YM}
{\cal L}_{\rm YM} = {\rm Tr}(B^+ F) - g^2 {\rm Tr}(B^+ B^+).
\ee
Here $B^+$ is a Lie algebra valued {\it self-dual} two-form field, and $F$ is the field strength for the gauge field. The unspecified contraction of the $B^+$ and $F$, and of the $B^+$ with itself in the second term is either (a multiple of) the wedge product, or the metric contraction (the two obviously coincide for the self-dual objects). Integrating out the two-form field one obtains the Lagrangian of YM in the more familiar form $(1/4g^2){\rm Tr}(F^+ F^+)$, which is nothing but the standard Lagrangian plus a topological term. Here $F^+$ is the self-dual part of the field strength.

As one sends $g^2\to 0$, the resulting theory is self-dual YM, which describes fields satisfying $F^+=0$. These are purely anti-self-dual gauge-fields, or YM instantons. The Lagrangian (\ref{YM}), as well as its self-dual limit $g^2=0$ has played in important role in various developments, see below. It is very convenient to have a viewpoint on YM as the self-dual theory corrected by a certain additional quadratic interaction.

Up to now, there was no analogous point of view on General Relativity. The new description (\ref{L}) proposed here achieves exactly this. Indeed, as we are going to demonstrate below, the theory (\ref{L}) with $\lambda=0$ describes (anti-) self-dual gravity. Adding the simple quadratic in the $B$ field term converts the (anti-) self-dual gravity into full GR, precisely paralleling what happens in the YM case described above. 

The main difference with the YM case (\ref{YM}) is that the Lagrangian (\ref{L}) is much more complicated than (\ref{YM}). In particular, it is non-polynomial in the fields, unlike (\ref{YM}) that is cubic. However, it could have been anticipated from the start that there cannot be any simple BF-type polynomial in the fields description of GR. Indeed, in the case of YM the non-triviality of (\ref{YM}) with $g^2=0$ consists in the fact that $B^+$ is required to be self-dual from the very start. This is possible, because in the YM case there is the metric on the underlying manifold, and this metric can be used to define what $B^+$ means. In the case of gravity there is no underlying metric, and the terms we can add to the Lagrangian should be constructed from the basic fields -- two-forms $F$ and $B$ -- in the covariant way. Anything simple and polynomial in these basic fields results in the topological theory. This is why the non-polynomial term is needed to obtain something non-trivial. 

Towards the end of this letter we give more comments explaining why (\ref{L}) may be important.  Let us now explain proofs of all the statements made above. To understand the constructions below, we need to remind the reader that (in an appropriate sense non-degenerate) triple of two-forms defines a conformal metric. Geometrically, this is the unique conformal metric in which all the given 3 two-forms are self-dual. An explicit expression for this conformal metric is due to Urbantke \cite{Urbantke}, but we will not need it here. A metric in the conformal class defined by the conformal metric is obtained by fixing a volume form. Let us now state our assertions more precisely. 

\begin{proposition} Let $\alpha,\lambda\not=0$ and $\lambda\not=3\alpha$. Consider critical points of the action functional with (\ref{L}) as the Lagrangian. The metric obtained by declaring the two-forms $B^i$ self-dual and with the volume form given by
\be\label{vol}
({\rm vol}) = \left({\rm Tr}(\sqrt{B^i\wedge B^j}) \right)^2
\ee
is Einstein with non-zero cosmological constant
\be\label{Lambda}
\Lambda= \lambda-3\alpha.
\ee
\end{proposition}
\begin{proposition} Let $\alpha\not=0,\lambda=0$. Consider critical points of the action functional with (\ref{L}) as the Lagrangian. The metric obtained by declaring the two-forms $B^i$ self-dual and with the volume form given by (\ref{vol}) is anti-self-dual Einstein with the cosmological constant $\Lambda= -3\alpha$.
\end{proposition}
Both propositions stated hold locally, in a region where no eigenvalues of the matrix $B^i\wedge B^j$ degenerate and one and the same branch of the matrix square root is taken at all points of the region.

Both assertions are proven by considering the equation resulting by varying (\ref{L}) with respect to the two-form field. We get
\be\label{FB}
F^i = \lambda B^i - \alpha {\rm Tr}(\sqrt{X_B}) (X_B^{-1/2})^{ij} B^j,
\ee
where we have defined the matrix 
\be
X_B^{ij} \sim B^i\wedge B^j.
\ee
The proportionality here is to be understood in the sense of the matrix-valued 4-form being divided by an arbitrarily chosen volume form (orientation). 

We now want to invert (\ref{FB}). This is achieved by
\be\label{BF}
B^i = \frac{1}{\lambda} \left( \frac{\alpha}{\lambda-3\alpha} {\rm Tr}(\sqrt{X_A}) (X_A^{-1/2})^{ij} + \delta^{ij}\right) F^j,
\ee
where now
\be
X^{ij}_A \sim F^i\wedge F^j.
\ee
We see already at this stage that when $\lambda=0$ the relation (\ref{FB}) cannot be solved for $B$ in terms of $F$. So, this case has to be considered separately. Let us continue with our analysis of the $\lambda\not=0$ case. Another useful relation is between the matrices of the square roots of the wedge products. We have
\be
(X_B^{1/2})^{ij} = \frac{1}{\lambda}\left( \frac{\alpha}{\lambda-3\alpha} {\rm Tr}(\sqrt{X_A}) \delta^{ij} + (X_A^{1/2})^{ij} \right),
\ee
from which we get
\be\label{traces}
{\rm Tr}(\sqrt{X_A}) = (\lambda - 3\alpha) {\rm Tr}(\sqrt{X_B}).
\ee

Let us now consider the other field equation, namely the one obtained by varying (\ref{L}) with respect to $A$. This equation reads simply
\be\label{feq2}
d_A B^i = 0.
\ee
Now, given the expression (\ref{BF}) for $B$ in terms of $F$, and the assumption $\alpha,\lambda\not=0$, we see that the field equation (\ref{feq2}) implies that
\be\label{pleb1}
d_A \Sigma^i=0,
\ee
where we have defined
\be\label{pleb2}
\Sigma^i := \frac{1}{\lambda-3\alpha} {\rm Tr}(\sqrt{X_A}) (X_A^{-1/2})^{ij} F^j .
\ee
Indeed, the term in $B$ proportional to $F$ can be dropped because it automatically satisfies $d_A F^i=0$ as its Bianchi identity. 

We now have, in view of the definition of $\Sigma$'s and the relation (\ref{traces})
\begin{gather}\label{pleb3}
\Sigma^i \wedge \Sigma^j \sim \delta^{ij}, \\ \label{pleb4} 
\Sigma^i\wedge \Sigma^i = \left({\rm Tr}(\sqrt{B^i\wedge B^j}) \right)^2.
\end{gather}
Equations (\ref{pleb1})-(\ref{pleb4}) taken together imply the following. Equation (\ref{pleb3}) is the Plebanski "metricity" condition that implies that $\Sigma$'s can be taken as an orhtonormal basis in the space of self-dual two-forms of some metric. This metric is then uniquely determined by $\Sigma$'s, and its volume form is $\Sigma^i\wedge \Sigma^i$. The relation (\ref{pleb4}) then implies that this volume form is that same as in the statement of the Proposition 1. Now, since $\Sigma$'s are declared self-dual, it is clear that $F$'s are self-dual, and hence the $B$'s are self-dual as well. This means that the metric obtained by declaring the $B$'s self-dual, with the volume form as in the statement of Proposition 1 is the same metric as defined by $\Sigma$'s. 

But now equations (\ref{pleb1}) imply that $A$ is the self-dual part of the Levi-Civita connection for the metric defined by $\Sigma$'s (and therefore by $B$'s). Finally, equation (\ref{pleb2}) implies that the curvature of the self-dual part of the Levi-Civita connection is self-dual as a two-form. This is the Einstein condition. We can also read off the cosmological constant. It is clear that it is given by (\ref{Lambda}). The Proposition 1 is proved.

Let us now see what changes in the case $\lambda=0$. In this case we cannot invert the relation (\ref{FB}). However, it is easy to see that in this case (\ref{FB}) implies 
\be\label{inst}
F^i \wedge F^j \sim \delta^{ij}.
\ee
We therefore define in this case
\be\label{inst2}
\Sigma^i := -\frac{1}{3\alpha} F^i.
\ee
In view of (\ref{inst}) and (\ref{traces}) we again have both of the relations in (\ref{pleb3}). We now have a metric defined by $\Sigma$'s, and since $\Sigma$'s are multiples of $F$'s, they satisfy the equations (\ref{pleb1}) as the consequence of the Bianchi identity. Therefore, again the connection $A$ receives the interpretation of the self-dual part of the Levi-Civita connection of the metric. The definition (\ref{inst2}) of $\Sigma$'s then states that the curvature of this connection is self-dual as the two-form. Moreover, there is no self-dual Weyl part of the curvature. This means that the metric is anti-self-dual Einstein, with cosmological constant as in the statement of Proposition 2. Hence, this proposition is proved. 

In the case $\lambda=0$ we have not used the field equation $d_A B^i=0$ in any way. From (\ref{FB}) we know that the two-forms $B^i$ are self-dual with respect to the metric defined by $\Sigma$'s. The equation $d_A B^i=0$ then simply states that the self-dual two-forms $B^i$ are covariantly constant with respect to the spin connection. 

For readers familiar with the "pure connection" formulation \cite{Krasnov:2011pp} of GR, we now give another proof of the above Propositions. It consists in simply integrating out the two-form field from (\ref{L}) and passing to the pure connection formulation. Most of the steps required for this procedure have already been carried out above. So, we simply substitute the $B$'s from (\ref{BF}) into (\ref{L}) and obtain the following Lagrangian
\begin{multline}\label{pure-con}
{\cal L}_{\rm GR}[A] \\ = \frac{\alpha}{2\lambda(\lambda-3\alpha)} \left({\rm Tr}(\sqrt{F^i\wedge F^j}) \right)^2 + \frac{1}{2\lambda} F^i\wedge F^i.
\end{multline}
The second term here is clearly just the topological term that does not affect field equations. The first term is the Lagrangian of GR in the pure connection formulation \cite{Krasnov:2011pp}. This gives a convincing demonstration that the original Lagrangian (\ref{L}) describes GR. The coefficient appearing in front of the first term in (\ref{pure-con}) explains why all of the assumptions $\alpha,\lambda\not=0$ and $\lambda\not=3\alpha$ are needed to reproduce General Relativity.

It is also interesting to see what happens in the $\lambda\to 0$ limit with the pure connection description (\ref{pure-con}). Let us rewrite the Lagrangian in terms of the eigenvalues of the matrix 
$X_A = {\rm diag}(\lambda_1, \lambda_2,\lambda_3)$. We get
\begin{widetext}\begin{multline}
{\mathcal L}_{\rm GR}[X_A]\Big|_{\lambda\to 0} \to \frac{1}{2\lambda} \left( \lambda_1+\lambda_2+\lambda_3- \frac{1}{3} (\sqrt{\lambda_1}+\sqrt{\lambda_2}+\sqrt{\lambda_3})^2\right)
 \\ = \frac{1}{6\lambda}\left(
(\sqrt{\lambda_1}-\sqrt{\lambda_2})^2 +(\sqrt{\lambda_2}-\sqrt{\lambda_3})^2 +(\sqrt{\lambda_3}-\sqrt{\lambda_1})^2 \right).
\end{multline}
\end{widetext}
This clearly shows that as $\lambda\to 0$ the Lagrangian blows up while at the same time making all 3 eingenvalues of the matrix $X_A$ coincide. It is then known that the condition that $X_A$ is proportional to the identity matrix is the instanton condition in the pure connection language. 

We conclude by giving some more remarks on why the proposed here formulation of GR may be useful. As we have already explained above, its main feature is that the full GR is obtained from what describes its (anti-) self-dual sector by just appending a simple quadratic in the two-form field term to the Lagrangian. The point is then that the ASD sector of General Relativity is integrable. At a formal level, this means that one can write down the general solution of ASD Einstein equations. Or, put it differently, one can do a change of variables that map the ASD field equations into "free" ones. In the context of Penrose's twistor description the ASD equations are interpreted \cite{Penrose:1976js} as a version of Cauchy-Riemann equations guaranteeing integrability of a certain complex structure on the twistor space. Thus, if ASD gravity described by the first two terms in our Lagrangian (\ref{L}) admits a simple twistor description, the it is perhaps possible to add one more term to the simple twistor ASD gravity Lagrangian and obtain a twistor description of the full GR. This would achieve what Penrose's twistor program \cite{Penrose:1977in} envisaged but never realised. We also note that twistor Lagrangians for ASD gravity have already been constructed in the literature see \cite{Wolf:2007tx}, \cite{Mason:2007ct}.

It is also worth noting that the same program in the context of Yang-Mills theory has to a large degree been completed. Thus, a twistor space description of the self-dual Yang-Mills theory has been given in \cite{Witten:2003nn}, and further developed in \cite{Mason:2005zm}. Both papers also provide a description of the twistor action for the full YM, incorporating the twistor space version of the second term in (\ref{YM}). An important further insight was given in \cite{Boels:2007qn}, deriving the so-called MHV rules \cite{Cachazo:2004kj} from the YM twistor action. The new Lagrangian description of GR proposed in this paper paves the way to realising the same program in the context of gravity.

Our work is also of relevance for the spin foam approach to quantum gravity \cite{Perez:2012wv}. The spin foam description of the topological theory, whose Lagrangian is (\ref{L}) with $\alpha=0$, is considered to be understood. Thus, if it was possible to give a spin foam description of the ASD gravity (\ref{L}) with $\lambda=0$, then it would perhaps be also possible to combine the two and obtain full GR. Given that the theory with $\alpha=0$ is believed to give rise to the quantum group ${\rm SU}_q(2)$, our description thus points in the direction of full GR being about "q-deformed instantons", whatever that may be. 

YH was supported by ENS Lyon and KK was supported by ERC Starting Grant 277570-DIGT.

\end{document}